\documentclass[pre,floatfix,showpacs,twocolumn]{revtex4}
\usepackage{amsmath}
\usepackage{amssymb}
\usepackage{graphicx}

\begin{document}

\title{Aggregates of rod-coil diblock copolymers adsorbed at a surface}

\author{C. Nowak, T.A. Vilgis}

\affiliation{Max-Planck-Institut f\"ur Polymerforschung, 
             Ackermannweg 10, 
             55128 Mainz, 
             Germany}


\begin{abstract}
  The behaviour of rod-coil diblock copolymers close to a
  surface is discussed by using extended scaling methods.  The copolymers are
  immersed in selective solvent such that the rods are likely to aggregate to
  gain energy. The rods are assumed to align only parallel to each other, such
  that they gain a maximum energy by forming liquid crystalline 
  structures.  If an aggregate of these copolymers adsorbs with the rods
  parallel to the surface the rods shift with respect to each other to allow
  for the chains to gain entropy.  It is shown that this shift decays with
  increasing distance from the surface.  The profile of this decay away from
  the surface is calculated by minimisation of the total free energy of the
  system.  The stability of such an adsorbed aggregate and other possible
  configurations are discussed as well.
\end{abstract}

\pacs{82.35.Gh Polymers on surfaces;
      82.35.Jk Copolymers, phase transitions, structure;  
      36.20.Ey Conformation (statistics and dynamics)}

\maketitle

\section{Introduction}

Rod-coil copolymers in selective solvents show a rich phase behaviour.
Depending on the chain length and the solvent quality they may form
cylindrical micelles or lamellar sheets. Similar phases can be found in melts
of rod-coil copolymers. These systems are therefore widely studied in the
literature
\cite{dowell,semenov1,halperin2,halperin1,vilgis1,williams1,holyst1,holyst2,matsen,friedel}.
Less extensive are the studies on a single multiblock polymer composed of
stiff rods which are connected by flexible chain spacers, see
\cite{grosberg,semenov2,nowak1,nowak2}. It was shown, that these polymers can
also form micellar and multi-micellar structures. The structural
behaviour of dissolved rod-coil copolymers in the presence of a surface is far
less understood. Rod-coil polymers grafted to a repulsive surface are shown 
to form 'turnip'- or 'jellyfish'-like micelles on top of the surface \cite{sevick}. 
However, to our knowledge there exists no study of rod-coil polymers in the 
presence of an attractive surface. 

In this paper we consider rod-coil diblock copolymers in selective solvent
close to a surface which is highly attractive for the rods and neutral to the
flexible parts of the copolymer.  Further, the solvent is assumed to be poor
for the rods, such that they align and tend to form aggregates,
and good for the chains. 
In addition it is assumed that the rods have a certain chemical
modification, such that they prefer to be parallel oriented with respect to
each rather than antiparallel. The aggregation behaviour of such rod-coil
copolymers, showing parallel alignment of the rods only, has been
investigated experimentally and computationally, see
\cite{stupp1,stupp2,sayar}.  We assume the energy penalty for antiparallel
alignment of two rods to be much higher than the energy penalty for these rods
being fully exposed to the solvent. In aggregates of these copolymers the
flexible parts therefore stick out in one direction only, see
Fig.(\ref{detach-fig}). If such an aggregate adsorbs with the rods parallel to
the surface, the rods shift with respect to each other to allow for the chains
to gain entropy, see Fig.(\ref{shift-fig}). The nature of this shift will be
examined in the following.

For simplicity we consider a quasi two-dimensional system.  That means the
width of the system in $y$-direction is equal to the rod diameter $d$. This
system can be viewed as a narrow slice of a system with infinite extension in
$y$-direction. Each of the diblock copolymers under consideration is composed of
a stiff rod of length $L$ and diameter $d$ to which a fully flexible chain of
$N$ monomers with monomer size $b$ is grafted. The solvent is characterised by
an energy penalty $\gamma$ per unit area of a rod exposed to the solvent. 
The energy gain $-\kappa$ per unit area of a rod for being in contact with the surface 
has to be chosen such that an aggregate actually adsorbs to the surface without 
dissociating into single rod-coil copolymers. 

The paper is organised as follows. In a first naive approach one would assume
a constant shift of the rods with respect to each other. This assumption leads to
an artefact in the shifting behaviour, as we show in the appendix. The
calculations in the appendix are presented in some detail because some of the
intermediate results are used in section II and III.  In a more sophisticated
approach the shift is allowed to vary with distance from the surface such that
it can develop a profile. This approach to the problem is presented in
section II. Some remarks on the stability of the adsorbed structure and the 
corresponding range of values for $\kappa$ are made
in section III. In section IV we finish with a brief discussion of the
results of the foregoing sections.

\section{Profile of the shift}

When an aggregate of rod-coil copolymers becomes adsorbed by a strongly attractive surface, an
additional confinement for the corona of the free chains may introduce new
effects. One possibility is that the entropy penalty of the confinement of
the corona chains due to the wedge defined by the surface and the rod aggregate 
prevents adsorption. Another extreme effect can be the destruction of the
aggregate. In between, the balance between entropy and energy can be
such, that the rods are shifted with respect to each other to allow for the chains to 
gain entropy. If this shift does not get to large, the rods stay together.
Such conformations appear possible whenever the defect
energy is balanced by the gain of entropy due to reduction of the wedge confinement.

The most naive assumption that the shift may be constant, i.e., it forms a
line with a certain slope, turns out to be unphysical in many
respects (see Appendix). Therefore it is necessary to introduce a curved deviation 
of the shift for each successive rod, which results from a total balance
of all entropic and enthalpic contributions.

\begin{figure}
\includegraphics[width=\linewidth]{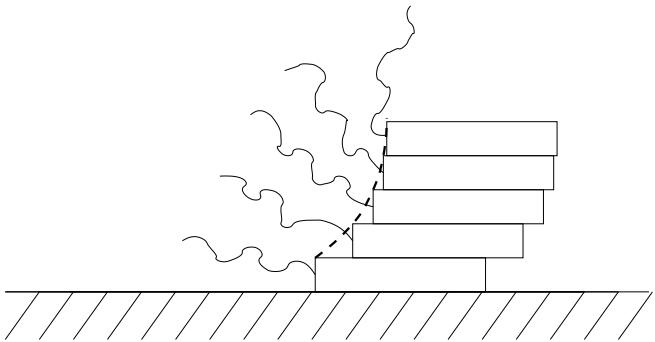}
\caption{\label{shift-fig}Aggregate of rod-coil copolymers adsorbed at a 
  surface. The shift of the rods with respect to each other decays with
  increasing distance from the surface.}
\end{figure}

We present in  this section the basic model which allows us to
calculate the shift of the rods as a function of distance from the surface.  From
intuition we expect the rods close to the surface to shift more than the rods
further away from the surface, since for the corresponding chains close to the
surface there is more entropy to gain than for the ones further away.
Therefore we expect an equilibrium conformation similar to the one shown in
Fig.(\ref{shift-fig}).

\begin{figure}[h]
\includegraphics[width=\linewidth]{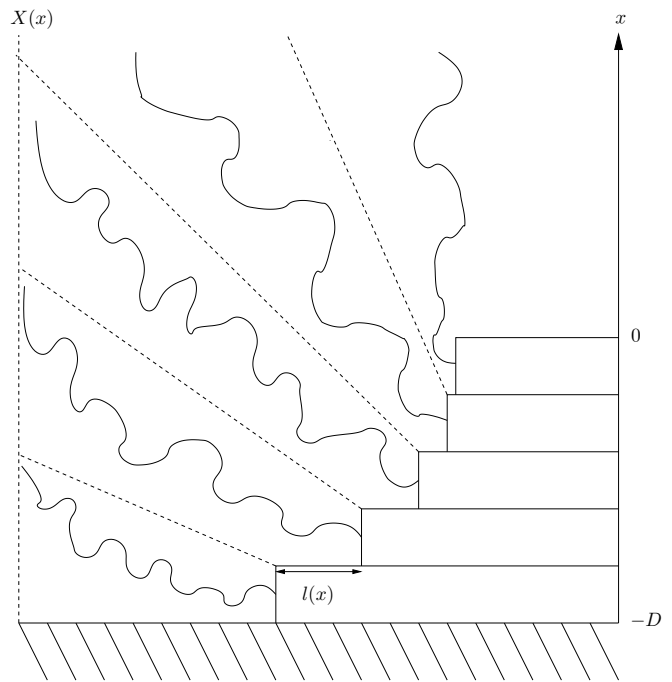}
\caption{\label{shift-splay-fig} This sketch shows the spatial segments or
  boxes filled by each chain. They get larger with increasing distance from
  the surface.  The shift at each position $x$ is denoted by $l(x)$. $x$
  ranges from $-D$ at the surface to 0 at the last rod.  The splay of the
  chains is given by $u(x)=X(x)-x$. }
\end{figure}

The dashed line in Fig.(\ref{shift-fig}) can be interpreted as the profile of
the shift $l$ as a function of the distance $x$ from the surface. The chains
are described by a local Flory-type model similar to the one used in
\cite{vilgis2} to describe a finite polymer brush.  In this model the free
energy of the system is given by the sum of an elastic term and an excluded
volume term for the chains plus a term which quantifies the energy penalty for
the additional rod surface exposed to the solvent due to the shift. We
construct the free energy such that it is a function of the splay of the
chains $u(x)$ - see Fig.(\ref{shift-splay-fig}) - and the shift of the rods
$l(x)$. At all positions $x$ we can safely assume $l(x)$ to be small compared to the
height $h$ of the brush-like structure formed by the chains. If the shift $l$ would be 
of the same order as $h$, the chains would hardly see each other, so there would be no 
driving force for a shift.  
Hence we can
assume $h$ to be constant for all $x$. 
The height is given by the equilibrium
height of a polymer brush in good solvent, i.e. $h = 4^{-1/3} b\,N
(b/d)^{2/3}$. The excluded volume parameter $v$ is set to $v=b^3$ for
simplicity.

\begin{figure}[h]
\includegraphics[width=\linewidth]{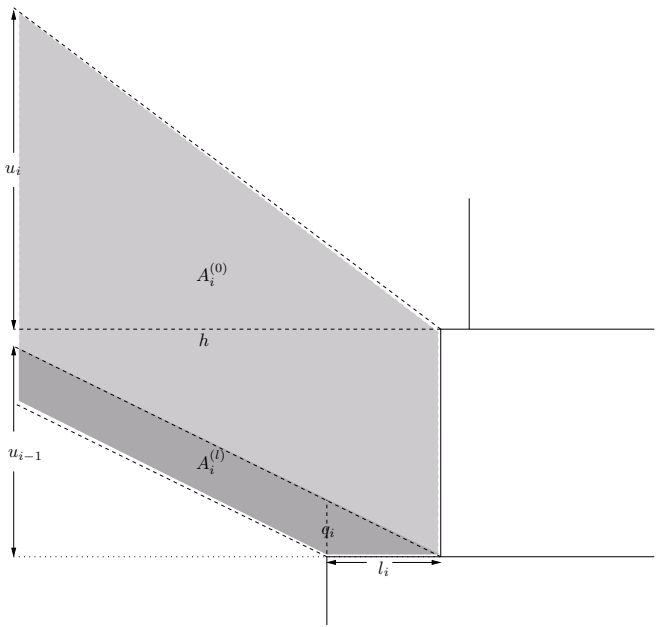}      
\caption{\label{vol-fig}This figure shows how the volume occupied by each chain 
is calculated as a function of splay $u$ and shift $l$.}
\end{figure}

Fig.(\ref{shift-splay-fig}) helps to define and explain how the excluded
volume term of the free energy for the shift geometry is constructed. 
Note, that it is only a sketch. The ratio of shift $l$ to brush height  
is much smaller than depicted in Fig.(\ref{shift-splay-fig}).
A chain starts at the rod and ends at the line $X(x)$ shown in
Fig.(\ref{shift-splay-fig}). It is assumed to fill the volume of the box given
by the dashed lines around the chains. We are aware that this assumption is
not valid for the chains far away from the surface but for these chains the
contribution of the excluded volume term is certainly very small. Hence the
assumption does not affect the total free energy in a significant way. The
splay $u(x)$ is given by $X(x)-x$.  To explain how to calculate the volume
available to each chain, Fig.(\ref{vol-fig}) shows a larger sketch of one of
the dashed boxes surrounding each chain in Fig.(\ref{shift-splay-fig}).

Inasmuch as we are going to consider a quasi two-dimensional system only, the
volume is given by the grey area in Fig.(\ref{vol-fig}) times the diameter of
the rods $d$.  The area $A_i^{(0)}$ shaded in light grey is given by $h
\left(d+\Delta u_i/2\right)$, where $\Delta u_i = u_i-u_{i-1}$. It is the area
$A^{(l)}_i$ shaded in dark grey where the shift of the rods $l(x)$ comes into
play. It is given by
\begin{equation}
A^{(l)}_i = h q_i - \frac{1}{2} l_i q_i.
\end{equation}
As can be seen from Fig.(\ref{vol-fig}) the length $q_i$ is given by 
\begin{eqnarray}
q_i = l_i \frac{u_{i-1}}{h} = l_i \frac{u_i-\Delta u_i}{h}\\
\label{Ai}
\Rightarrow A^{(l)}_i = l_i\left(u_i-\Delta u_i\right)\cdot\left(1-\frac{l_i}{2 h}\right).
\end{eqnarray}
Since we assumed $l_i$ to be small compared to $h$, the term in the last
brackets in Eq.(\ref{Ai}) can be approximated by $1$. In Fig.(\ref{vol-fig})
this corresponds to double counting the area of the small triangle with the
catheti $q_i$ and $l_i$.

In order to construct a free energy functional that could be minimised with
respect to the shift and splay shapes $l(x)$ and $u(x)$ we take the continuum
limit: $l_i \rightarrow l(x)$, $u_i \rightarrow u(x)$, $\Delta u_i \rightarrow
d u'(x)$.  Taking the continuum limit and adding the two contributions, the
total volume $V(x)$ available to a chain at position $x$ is given by
\begin{equation}
\label{vol}
V(x) = d^2 h \left(1+u'(x)/2\right) + d l(x) (u(x)-d u'(x)).
\end{equation}  
The elastic term of the free energy is given by a contribution proportional to
$h^2$ representing the stretching of the chains away from the rods and a
contribution proportional to $u(x)^2$ representing the stretching of the
chains parallel to the rods.  The energy penalty for the additional area of a
rod at position $i$ exposed to the solvent as a function of the shift $l_i$ is
simply given by $2 \gamma l_i d$.  The complete free energy functional can now
be constructed.
\begin{eqnarray}
\label{Fdim}
\beta F = 2 \int\limits_{-D}^{0} \mathrm{d}x \beta \gamma l(x)
+ \frac{1}{N b^2 d}\int\limits_{-D}^{0} \mathrm{d}x \left[ u(x)^2+h^2 \right]\nonumber\\
+ \frac{N^2 b^3}{2 d} \int\limits_{-D}^{0}\mathrm{d}x \left[\frac{1}{h d^2(1+u'(x)/2)+d l(x)(u(x)-d u'(x))}\right]
\nonumber\\
\end{eqnarray}
For
simplicity we dropped all terms which do not depend on either $l$ or $u$. 
As already mentioned the excluded volume parameter $v$ is set to $v=b^3$. 

The excluded volume term in Eq.(\ref{Fdim}) was constructed assuming constant 
density of monomers for each chain within the box of volume $V$ (Eq.(\ref{vol})). 
The monomer density close to the rods is larger than the one further away from the rods 
and therefore this assumption tends to underestimate the excluded volume energy. 
However, it is the standard approximation used in Flory-type models and has been 
proven to be sufficient to describe the behaviour of a finite polymer brush, 
see \cite{vilgis2}.

The total length $D$ of the rod aggregate perpendicular to the surface is 
assumed to be $D \ge h$. This is explicitly needed in the appendix.

Fig.(\ref{shift-splay-fig}) shows that the rod at the surface has zero shift
since there is no other rod underneath with respect to which it could shift.
So for the rod-coil copolymer at the surface the integrand in Eq.(\ref{Fdim})
reduces to the one in Eq.(\ref{Fbrush}) in the appendix (with $\sigma=1/d^2$).
This of course also means that the equation for the shift (Eq.(\ref{l}) below)
is only valid from the second rod on (as counted from the surface).
   
To calculate the equilibrium shift $l(x)$ it is necessary to compute the
Euler-Lagrange equations from a functional minimisation of Eq.(\ref{Fdim})
with respect to $u(x)$ and $l(x)$.  The Euler-Lagrange equation for the splay
$u(x)$ has the first integral
\begin{eqnarray}
\label{firstIdim}
2 \beta \gamma l(x) + \frac{u(x)^2 + h^2}{N b^2 d}\nonumber\\ 
+\frac{N^2 b^3}{d}\left[\frac{1}{h d^2(2+u'(x))+2dl(x)(u(x)-d u'(x))}\right.\nonumber\\
+\left.\frac{\left(h-2l(x)\right)u'(x)}
{\left(hd(2+u'(x))+2l(x)(u(x)-d u'(x))\right)^2}\right] = C_1.
\end{eqnarray}
Variation of the free energy functional in Eq.(\ref{Fdim}) with respect to $l(x)$ yields
\begin{equation}
\label{ELl}
2 \beta \gamma  + \frac{2 N^2 b^3}{d^2}\left[\frac{u(x)-d u'(x)}{\left(hd(2+u'(x))+2l(x)(u(x)-d u'(x))\right)^2}\right] = 0.
\end{equation}
The quadratic Eq.(\ref{ELl}) can be solved for the shift $l(x)$. 
\begin{equation}
\label{l}
l(x) = \frac{\left(\frac{b^3 N^2}{d^2\beta\gamma}(u(x)-d u'(x))\right)^{\frac{1}{2}}-hd(2+u'(x))}{2(u-du'(x))}
\end{equation}
Inserting the expression for the shift - i.e. Eq.(\ref{l}) - into
Eq.(\ref{firstIdim}) results in a complicated, highly nonlinear differential
equation which cannot be solved exactly. However, the overall effect of the
shift on the free energy is certainly smaller than the overall effect of the
splay. Thus it is a reasonable approximation to calculate the solution for the
splay at zero shift and to use this as an approximation for the splay $u$ in
Eq.(\ref{l}).  We are aware that this approximation breaks down when the splay
becomes very small close to the surface.  Nevertheless, for zero splay the
shift should be constant.  The shape of the profile of the shift away from the
surface at finite splay can therefore be calculated within this approximation.

\begin{figure}[h]
\includegraphics[width=\linewidth]{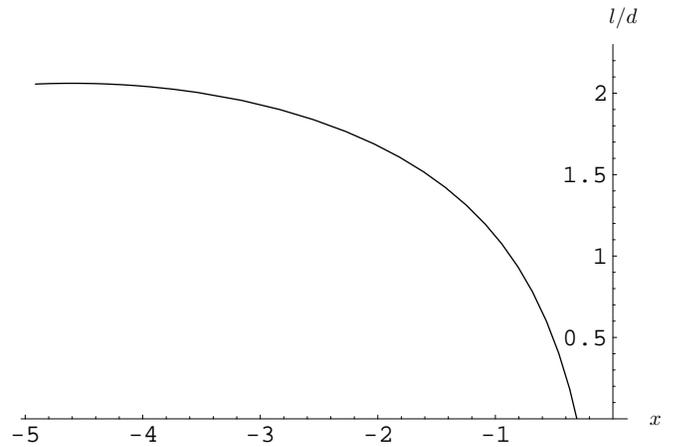}      
\caption{\label{l-x-paramplot} The shift $l$ is plotted as a function of negative distance. 
  The rod furthest from the surface 
  is located at $x=0$, compare Fig.(\ref{shift-splay-fig}). Parameters:
  $N=700, b/d=0.1, \beta\gamma d^2=1$}
\end{figure}

Now we calculate an approximate solution of the splay $u$ for zero $l$. If the
shift $l(x)$ in Eq.(\ref{firstIdim}) is set identical zero, the differential
equation equation reduces to
\begin{eqnarray}
\frac{u(x)^2 + h^2}{N b^2 d} 
+\frac{2 N^2 b^3}{hd^3}\left[\frac{1+u'(x)}{(2+u'(x))^2}\right] = C_1\\
\label{firstI-l0}
\Rightarrow u^2 + 8 h^2\frac{1+u'}{(2+u')^2} = C_2
\end{eqnarray}
which is Eq.(\ref{1_int}) from the appendix as should be.  It is convenient to
introduce dimensionless variables $\tilde{u}=u/(\sqrt{8}h)$, 
$\tilde{x}=x/(\sqrt{8}h)$ and $\tilde{C}_2=C_2/(8 h^2)$.  Eq.(\ref{firstI-l0})
can be integrated for arbitrary $\tilde{C}_2$ which leads to an implicit
equation for the splay $\tilde{u}(\tilde{x})$ similar to
Eq.(\ref{implicit_uD})
\begin{eqnarray}
\label{imp_u0-prof}
2\tilde{u}_0 - 2\tilde{u} + \ln \left[2 \tilde{u} +
(1-4\tilde{C}_2+4\tilde{u}^2)^{1/2}\right]\nonumber\\
-\ln \left[2 \tilde{u}_0 + (1-4\tilde{C}_2+4\tilde{u}_0^2)^{1/2}\right] = 4\tilde{x}.
\end{eqnarray}
Note that $-\tilde{D} \le \tilde{x} \le 0$. The integration constant
$\tilde{C_2}$ can be determined by using the boundary condition
$\tilde{u}(-\tilde{D})=0$ - i.e. zero splay at the surface, see
Fig.(\ref{shift-splay-fig}).
\begin{equation}
\label{C2_u0} 
\tilde{C}_2=\frac{1}{4}\left(1-\tilde{u}_0^2 \sinh\left[2\tilde{D}+\tilde{u}_0\right]^{-2}+
\tilde{u}_0^2 \cosh\left[2\tilde{D}+\tilde{u}_0\right]^{-2}\right)
\end{equation}
The chain furthest away from the surface at $x=0$ can topple over completely and is therefore 
allowed a splay $u_0 = h$ or 
$\tilde{u}_0 = 1/\sqrt{8}$. Therewith Eqs.(\ref{imp_u0-prof},\ref{C2_u0}) reduce to
\begin{eqnarray} 
\label{imp-prof}
1/\sqrt{2} - 2\tilde{u} + \ln \left[2 \tilde{u} +
(1-4\tilde{C}_2+4\tilde{u}^2)^{1/2}\right]\nonumber\\
-\ln \left[1/\sqrt{2} + (3/2-4\tilde{C}_2)^{1/2}\right] = 4\tilde{x},\\
\label{C2}
\tilde{C}_2=\frac{1}{8}\left(2-\sinh\left[4\tilde{D}+1/\sqrt{2}\right]^{-2}\right).
\end{eqnarray}  
It is not possible at this stage to resolve Eq.(\ref{imp-prof}) with respect
to the splay $u$. However, we know $x$ as a function of $u$, $u'$ as a
function of $u$ (see Eq.(\ref{firstI-l0})) and $l$ as function of $u$ and
$u'$. Hence for a certain set of parameters, we can plot the shift $l$ as a
function of $x$ by either numerically resolving Eq.(\ref{imp-prof}) or by
showing a parameter plot of $l$ versus $x$, using $u$ as a parameter. This is
done for a characteristic set of parameters in Fig.(\ref{l-x-paramplot}). The
plot demonstrates that our assumption for the profile of the
splay - Fig.(\ref{shift-fig}) - was indeed reasonable. The shift increases
from the furthest rod towards the surface until it reaches its maximum value
in a reasonable form.

We are now going to estimate the threshold value of $\gamma$ above which 
the energy penalty for additional rod-solvent exposure becomes 
to large for  a shift to occur. 

The shift is identical zero if the right hand side of 
Eq.(\ref{l}) is less or equal to zero for all $x$.
\begin{eqnarray}
\frac{\left(\frac{b^3 N^2}{d^2\beta\gamma}(u(x)-d u'(x))\right)^{\frac{1}{2}}-hd(2+u'(x))}{2(u-du'(x))} 
\le 0
\nonumber\\
\Rightarrow \beta\gamma \ge \frac{b^3 N^2 (u(x)-d u'(x))}{h^2 d^4 (2+u'(x))^2}.
\end{eqnarray}
To find an upper limit for the threshold value $\gamma_c$ we note that 
the maximum value of the shift is $u_0 = h$. An upper estimate for 
$(u(x)-d u'(x))/(2+u'(x))^2$ is therefore given by $4 h$. The upper limit for 
$\gamma_c$ is thus given by
\begin{equation}
\beta\gamma_c \approx \frac{b^3 N^2}{4 h d^4} \approx 0.4\frac{N}{d^2}\left(\frac{b}{d}\right)^\frac{4}{3}.
\end{equation}
For the set of parameters chosen in Fig.(\ref{l-x-paramplot}) this yields the rough estimate of 
$\beta\gamma_c d^2 \approx 13$.

In this section an attempt was made to describe a variable shift $l(x)$. For
values of the splay $u(x)$ which are large enough to dominate the effect of
the shift, we found a set of equations (\ref{l}, \ref{imp-prof}, \ref{C2})
which determine $l$ as a function of $u$ and $u$ as a function of $x$.
Although it is not possible to resolve Eq.(\ref{imp-prof}) with respect to
$u$, the profile of the shift can be plotted for a certain set of parameters,
see Fig.(\ref{l-x-paramplot}). In this section we always assumed that the
adsorbed aggregate is internally stable and does not disintegrate in the sense 
that single copolymers leave the aggregate. Therefore we are
going to discuss in the next section under which conditions the adsorbed aggregate 
is stable and which other configurations are possible.
 
\section{Stability}  
In section II we assumed that an attached configuration of $f$ rod-coil copolymers at a
surface as it is pictured in Fig.(\ref{shift-fig}) is stable. There are three 
other possible configurations. 

\begin{figure}[h]
\includegraphics[width=\linewidth]{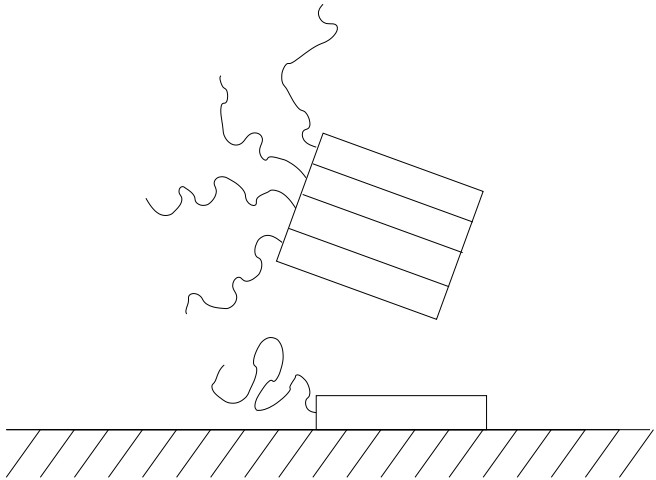}
\caption{\label{detach-fig} For long chains a detachment of the aggregate 
from the rod adsorbed to the surface might be preferable.}
\end{figure}

In this
quasi two-dimensional system only one rod is in contact with the surface.
Therefore a possible configuration is the one shown in
Fig.(\ref{detach-fig}), where one single copolymer is adsorbed at the surface
and the others form a detached sheet. We call such a situation detached
configuration in the following. The rods in the detached sheet might also 
prefer a shifted geometry. We refrain from a discussion of this shift, since this 
paper  is mainly concerned with the behaviour of a complete aggregate 
in contact with a surface. 

\begin{figure}[h]
\includegraphics[width=\linewidth]{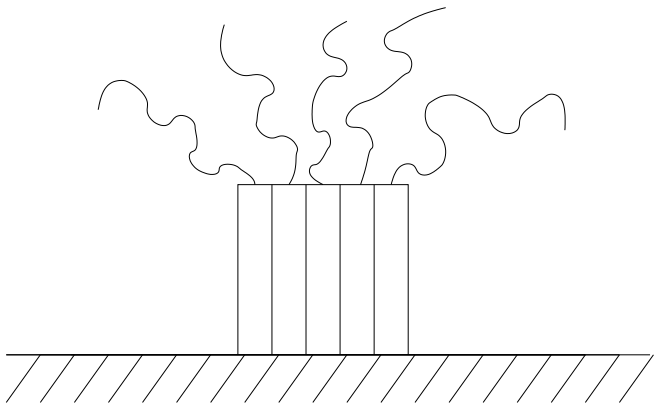}
\caption{\label{up-fig} Another possible configuration: The aggregated 
  rods adsorb perpendicular to the surface. This configuration is preferable
  for large aggregates.}
\end{figure}

The energy of the rod-surface contact in the detached configuration is the same as for the 
attached one.
A configuration with different contact energy is the mushroom-like one as depicted 
in Fig.(\ref{up-fig}). 
This configuration is always preferable to a complete detachment of the aggregate 
since in the latter case the system would gain no contact energy.
  
\begin{figure}[h]
\includegraphics[width=\linewidth]{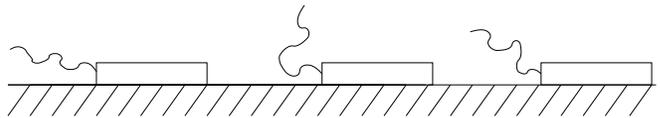}
\caption{\label{diss-fig} The aggregate dissociates and the individual rods 
adsorb at the surface. This configuration is preferred for large $\kappa$.}
\end{figure}

The last possible configuration is a complete dissociation of the aggregate into 
single copolymers due to the presence of the attractive surface. 
These single copolymers then adsorb individually at the surface, 
see Fig.(\ref{diss-fig}).
This configuration yields the highest gain of contact energy. However, it is also the 
configuration with the highest energy penalty for exposure of rod surface to the 
solvent. For very high contact energy, i.e. $\kappa \gg \gamma$, 
the system always dissociates. On the other hand, if $\kappa$ is to small the system might 
prefer the mushroom configuration even for long rods, since it allows for the chains 
to gain entropy without much increase in exposure of the rods to the solvent. 

By 
estimating the free energies of these configurations and comparing them with the one 
of the attached configuration, it is possible to find the range of $\kappa$ in which the attached 
configuration is stable. 
To achieve this at least approximatively we calculate the free energy
 of the attached configuration with zero shift. It gives a slight overestimation 
of the free energy of the configuration
considered in section II. However, we still get a rough estimate for the
parameter range in which the attached configuration is stable.  
At zero shift the chains form a finite brush. Its free energy is calculated 
in the appendix and given by Eq.(\ref{FbrushI}). One side of the brush is free 
and therefore allowed a splay of $u_0 = h$, the other side is confined by the 
surface, i.e. $u_D = 0$. The length $D$ is given by $fd$.
The free energy of the attached configuration then reads
\begin{equation}
\label{Fattach}
\beta F_{attach} = \frac{1}{N b^2 d}
\left[3fdh^2-\left(\frac{7}{6}-\frac{1}{\sqrt{2}}\right)h^3\right].
\end{equation}
We use this as a reference energy and add energy gains and penalties due 
to rod-surface contact or rod-solvent exposure to the free energies of the other 
configurations.

Compared to the attached configuration the contact energy of the mushroom 
differs by $\kappa(Ld-fd^2)$. The chains can also be described as a finite brush. 
Here both ends are free and are therefore allowed a splay of $u_0=u_D=h$. The free 
energy of the mushroom configuration is hence given by 
\begin{eqnarray}
\label{Fmushroom}
\beta F_{mushroom} &=& \kappa \left(dL - fd^2\right)\nonumber\\
&+&\frac{1}{N b^2 d}\left[3fdh^2-\left(\frac{7}{3}-\sqrt{2}\right)h^3\right].
\end{eqnarray}
The attached configuration is preferred to the mushroom configuration if
$F_{attach} < F_{mushroom}$. This yields the following condition for $\kappa$
\begin{equation}
\label{kappa-mush}
\beta\kappa > \frac{0.12 N^2}{Ld-fd^2}\left(\frac{b}{d}\right)^3.
\end{equation}
This is the lower bound for $\kappa$. To get the upper bound the free energy of the 
dissociated copolymers - see Fig.(\ref{diss-fig}) - has to be estimated. 

Compared to the attached configuration  the dissociated one yields a contact  energy gain 
of $-(f-1)\kappa Ld$. But on the other hand it also gives rise to an additional energy 
penalty of $2(f-1)\gamma Ld$.  Within this Flory-type theory the flexible chains of the 
individual copolymers at the surface can be treated as free ones and their free energy can
be neglected. Comparison of $F_{attach}$ and $2(f-1)\gamma Ld-(f-1)\kappa Ld$ 
yields the upper limit of $\kappa$ above which the system dissociates:
\begin{equation}
\label{kappa-diss}
\beta\kappa < 2\beta\gamma - \frac{1}{(f-1)Ld^2}\left[1.2f d N\left(\frac{b}{d}\right)^\frac{4}{3}
-0.12 N^2 b\left(\frac{b}{d}\right)^2\right]
\end{equation}
Within this range of values for $\kappa$ the attached configuration can actually be stable.
For the parameters chosen in Fig.(\ref{l-x-paramplot}), $N=700, b/d=0.1,
\beta\gamma d^2=1$ and $L/d=80$, $f=30$, this range is given by 
$1.18 < \beta\kappa d^2 < 1.52$.

It is also of interest to keep $\kappa$ and $\gamma$ fixed and to investigate at which combination 
of molecular properties of the copolymers which configuration is preferred. 
We focus here on the length of the rods $L$ and the number of chain 
monomers $N$.  
In the following the critical rod lengths which separate each two of the possible 
configurations from each other are calculated as functions of $N$ and of the other parameters. 
Since it depends on the specific combination of parameters, especially on $\gamma$ 
and $\kappa$, which configurations are neighbouring in $L$-$N$ space, all possible 
critical rod lengths are calculated. Note, that not all possible phase boundaries can exist 
for one given set of parameters.
For each one to exist, the set of parameters have 
to be chosen appropriately.  

Eq.(\ref{kappa-mush}) can be rearranged such that it gives the 
critical rod length below which the system changes form the attached to the 
mushroom configuration. 
\begin{equation}
L^{attach}_{mushroom} \approx  fd + 0.12\frac{N^2 b}{\beta\kappa d^2}\left(\frac{b}{d}\right)^2 
\end{equation}
Comparison of $F_{mushroom}$ with the energy of the dissociated configuration 
yields the rod length $L^{mushroom}_{dissociate}$ which separates the mushroom 
from the dissociated configuration. 
\begin{equation}
L^{mushroom}_{dissociate} \approx 
\frac{1.2 f d N \left(\frac{b}{d}\right)^{4/3}-0.24N^2 b\left(\frac{b}{d}\right)^2-fd \beta\kappa d^2}
{2(f-1)\beta\gamma d^2-f\beta\kappa d^2}
\end{equation}
In case of rather large $\kappa$ (close to the upper limit) there might exist a rod 
length which directly separates the attached configuration from the dissociated 
configuration. It is found to be 
\begin{equation}
L^{attach}_{dissociate} \approx \frac{1.2 f d N\left(\frac{b}{d}\right)^{4/3}-0.12N^2 b \left(\frac{b}{d}\right)^2}
{(f-1)(2\beta\gamma d^2-\beta\kappa d^2)}.
\end{equation}

To calculate the boundaries of the detached configuration (see Fig.(\ref{detach-fig})), its free energy has to be estimated. 
As already mentioned above, this configuration might also prefer a shifted geometry. But since we already 
estimated $F_{attach}$ for zero shift, is is sufficient to calculate also $F_{detach}$ for a rectangular sheet without shift. 
Compared to the attached configuration the free energy of the detached configuration has a contribution from two additional
rod surfaces exposed to the solvent, which is given by $2Ld\gamma$. 
The free energy of the chains of the detached sheet is similar to the one of the 
mushroom configuration,  with $D=f-1$. The free energy of the chain of the single 
copolymer can be neglected as in the case of dissociation. $F_{detach}$ is hence given 
by
\begin{equation}
\label{Fdetach}
\beta F_{detach} = 2Ld\beta\gamma
+\frac{1}{N b^2 d}\left[3(f-1)dh^2-\left(\frac{7}{3}-\sqrt{2}\right)h^3\right].
\end{equation} 
The rod length $L^{detach}_{attach}$ with separates the detached and the attached 
configuration can now be estimated.
\begin{equation}
L^{detach}_{attach} \approx  0.6\frac{N}{d\beta\gamma}\left(\frac{b}{d}\right)^{\frac{4}{3}} + 0.06\frac{N^2}{d\beta\gamma}\left(\frac{b}{d}\right)^3
\end{equation}
However, in $L$-$N$ space the detached configuration might sit in between the mushroom and 
the dissociated configuration. This is indeed the case for a wide range of parameters. 
Therefore equating Eq.(\ref{Fmushroom}) and Eq.(\ref{Fdetach}) gives the rod length 
which separates mushroom and detached configuration.
\begin{equation}
L^{mushroom}_{detach} \approx \left(fd-1.2\frac{N d}{\beta\kappa d^2}\left(\frac{b}{d}\right)^\frac{4}{3}\right)
\left(1-\frac{\gamma}{\kappa}\right)^{-1}
\end{equation}
The length $L^{detach}_{dissociate}$ which separates the detached from the 
dissociated configuration is found to be 
\begin{equation}
L^{detach}_{dissociate} \approx \frac{1.2 (f-1) d N\left(\frac{b}{d}\right)^{4/3}-0.24N^2 b \left(\frac{b}{d}\right)^2}
{2(f-2)\beta\gamma d^2-(f-1)\beta\kappa d^2)}.
\end{equation}
In the conclusions we use these critical rod lengths to calculate one phase diagram 
in $L$-$N$ space for a typical set of parameters as an example. 

In this section it was shown that the configuration pictured in Fig.(\ref{shift-fig}) 
can be stable within a certain range of values for $\kappa$. 
There are three other possible 
configurations as shown in Fig.(\ref{detach-fig}), Fig.(\ref{up-fig}) 
and Fig.(\ref{diss-fig}). 

\section{Conclusion}

A discussion of rod-coil copolymer aggregates adsorbed at a surface in a two
dimensional approximation was presented.  The aggregates form because of a
selective solvent, poor for the rods and good for the chains. Due to their
chemical structure the rods only align parallel to each other.  The surface is
assumed to be attractive for the rods and neutral with respect to the
chains.  If the aggregate adsorbs with the rods parallel to the surface, the
rods shift with respect to each other to allow for the chains to gain entropy
and to therefore lower their confinement energy. If the shift is assumed to be
constant for all rods, the system shows an artificial behaviour.  This can be
seen from the considerations in the appendix.  

In section II we constructed a
model which allows for the shift to vary. It was possible to partially solve
this model and to show that away from the surface the shift decays.  Close to
the surface the splay of the chains is close to zero and therefore the shift
is basically constant.  The region further away from the surface is the
interesting one showing the decay profile of the shift, see
Fig.(\ref{l-x-paramplot}).  

\begin{figure}[h]
  \includegraphics[width=\linewidth]{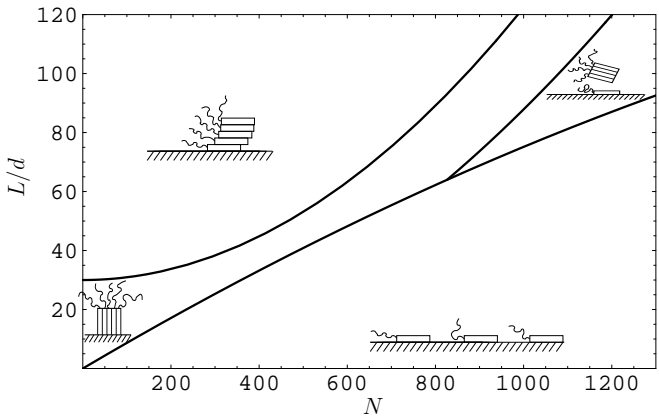}
\caption{\label{phasediag}Configurations of the rod-coil copolymers 
at the surface. Parameters: $b/d=0.1, f=30, \beta\gamma d^2=1, \beta\kappa d^2=1.3$}
\end{figure}

In section III we showed that the configuration
considered in section II can actually be stable within a certain range of contact 
energies between rods and surface. This range was calculated. 
Three other
possible configurations were also discussed, 
see Figs.(\ref{detach-fig},\ref{up-fig},\ref{diss-fig}).  
These considerations allow us to plot a phase diagram of the
configurations in $L$-$N$ space, which - for a typical set of parameters - is
shown in Fig.(\ref{phasediag}).
The contact energy per unit area $\kappa$ is chosen such that there exists 
a region in $L$-$N$ space in which the entropy loss of the confined chains is 
compensated by the energy gain due to rod-surface contact. But it is also 
chosen to be not much larger than the rod-rod contact energy $\gamma$, since 
otherwise the aggregate would dissociate. 
However, for very long chains the aggregate always dissociates into single copolymers 
which then individually adsorb. 
Nevertheless, Fig.(\ref{phasediag}) shows that there is indeed a broad region
in $L$-$N$ space in which the attached configuration is stable and the rods
shift with respect to each as discussed in section II.

For further considerations in the future it would be of interest to study the
behaviour of a finite three dimensional aggregate adsorbed with the rods
parallel to the surface.  We expect the profile of the shift to form a two
dimensional surface with the innermost rods close to the surface showing the
maximum shift.

\begin{appendix}
\section{Constant shift}
            
Despite the fact that the assumption of a constant shift appears unphysical and leads to
contradictory results we are discussing it in this appendix in some
detail. It is a useful example to define the arising problems in a clear way.

\begin{figure}[h]
\includegraphics[width=\linewidth]{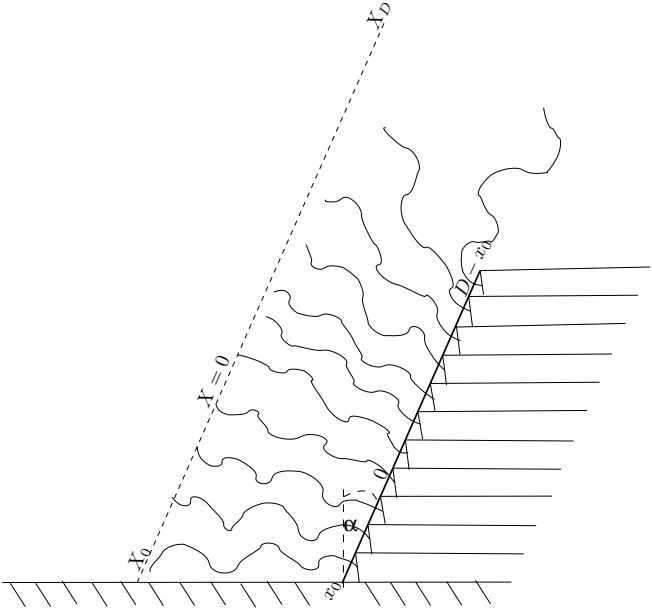}
\caption{\label{const-fig}Aggregate of rod-coil copolymers adsorbed at a surface. 
  The rods are all shifted with respect to each other by 
  the same distance. The shift is characterised by the angle $\alpha$. This
  sketch shows how we define the $x$-range and the splay. Note: $u(x)=X(x)-x$,
  i.e. $u_0=X_0-x_0$ and $u_D=X_D-(D-x_0)$.}
\end{figure}

The rods are assumed to shift a constant distance with respect to each other
as shown in Fig.(\ref{const-fig}). The characteristic quantity related to this
shift is the angle $\alpha$. This angle can be calculated by calculating the
free energy of the entire system and minimising it with respect to $\alpha$.

The additional free energy per rod due to the shift is given by the additional
surface of the rod exposed to the solvent $F_{rod} = 2 \gamma d^2 \tan\alpha$.
To calculate the free energy of the chains we treat them as if they would 
form a finite brush grafted to the surface shown as a thick line in
Fig.(\ref{const-fig}). For a finite brush the trajectories of the single
polymer chains are not all perpendicular to the grafting surface as for an
infinite one. The polymer chains show a splay $u$. Fig.(\ref{const-fig})
illustrates how this quantity is defined; that is, $u(x)=X(x)-x$.  The first
chain is allowed a splay of $u_0 = h \tan\alpha$ due to the surface, where $h$
is the brush height given by $h = 4^{-1/3} b\,N \sigma^{1/3} b^{2/3}$.  The
last chain can topple over completely and is therefore allowed a splay $u_L =
h$. The grafting density $\sigma$ is a function of $\alpha$ as well. It is
given by $\sigma(\alpha)=\cos(\alpha)/d^2$.  As in section II a Flory-type
approach is used to describe the free energy of the finite brush following the
lines of \cite{vilgis2}.  Each chain fills a box of volume
$h\sigma^{-1}(1+u'/2)$, with $u'=\mathrm{d}u/\mathrm{d}x$.  The free energy
for the finite brush is then given by
\begin{equation}
\label{Fbrush}
\beta F_{brush} = \frac{d}{N b^2}\sigma \left( \int\limits_{0}^{D} \mathrm{d}x \left[ u(x)^2+h^2 \right] 
+ 4 h^2 \int\limits_{0}^{D}\mathrm{d}x \left[\frac{1}{2+u'}\right] \right),
\end{equation}
where $D$ represents the total length of the brush. 
For our system of aggregates of rod-coil copolymers it is given by $D(\alpha)=f d \cos(\alpha)$.
The first integral of the Euler-Lagrange equation for the splay $u(x)$ obtained from Eq.(\ref{Fbrush}) 
is given by
\begin{equation}
\label{1_int}
u^2 + 8 h^2\frac{1+u'}{(2+u')^2} = C.
\end{equation}
Eq.(\ref{Fbrush}) does not distinguish between positive and negative splay.
Hence we have to separate our system into two finite brushes which meet at the
chain with zero splay, see Fig.(\ref{const-fig}).  This chain has to be
determined by an equilibrium condition (total length of the brush $D$).

It is convenient to introduce dimensionless variables $\tilde{u}=u/(\sqrt{8}h), \tilde{x}=x/(\sqrt{8}h)$ 
and $\tilde{C}=C/(8 h^2)$. 
Eq.(\ref{1_int}) can be integrated for arbitrary $\tilde{C}$ which leads to
\begin{eqnarray}
\label{implicit_uD}
\mathrm{I} &=& 
2 \tilde{u} - 2 \tilde{u}_D - \ln \left[2 \tilde{u} +
(1-4\tilde{C}+4\tilde{u}^2)^{1/2}\right]\nonumber\\
&+&\ln \left[2 \tilde{u}_D + (1-4\tilde{C}+4\tilde{u}_D^2)^{1/2}\right] 
= 4(\tilde{D}-\tilde{x}_0-\tilde{x}).
\nonumber\\
\end{eqnarray}
The integration constant can now be calculated by using the condition $2\mathrm{I}(\tilde{u}=0) -\mathrm{I}(\tilde{u}=\tilde{u}_0)$
\begin{eqnarray}
\tilde{C}&=&\frac{1}{4}\left(1-(\tilde{u}_0+\tilde{u}_D)^2 \sinh[2\tilde{D}+\tilde{u}_0+\tilde{u}_D]^{-2}\right.\nonumber\\
&+&\left.(\tilde{u}_0-\tilde{u}_D)^2 \cosh[2\tilde{D}+\tilde{u}_0+\tilde{u}_D]^{-2}\right).
\end{eqnarray}
The free energy in Eq.(\ref{Fbrush}) cannot be integrated directly using the implicit solution for the 
splay $\tilde{u}$, Eq.(\ref{implicit_uD}). Therefore we have to find an appropriate approximation. 
It can be shown that $1-4\tilde{C} \approx 0$ for all $u_0$ if $D \ge h$. 
Hence Eq.(\ref{implicit_uD}) can be very well approximated as 
\begin{eqnarray}
\label{u-approx}
2 \tilde{u} - 2 \tilde{u}_D - \ln \left[\frac{\tilde{u}}{\tilde{u}_D} \right] \tilde{u} = 4(\tilde{D}-\tilde{x}_0-\tilde{x}), 
\nonumber\\ 
\mathrm{for} \hspace{0.5cm} \tilde{u} \ge \tilde{u_c} = \frac{1}{2}(1-4\tilde{C})^{1/2}.
\end{eqnarray} 
For $0\le\tilde{u}\le\tilde{u_c}$ we can choose a linear approximation. 

To calculate the free energy we have to perform the following integration
\begin{equation}
\label{dimless-int}
\int \mathrm{d}\tilde{x} \left[ \tilde{u}^2+\frac{1}{2(2+\tilde{u}')} \right]. 
\end{equation}  
We first consider the regime $0\le\tilde{u}\le\tilde{u}_c$:
\begin{equation}
\tilde{u}'(\tilde{x}\le\tilde{x}_c)\ll 2 \Rightarrow \frac{1}{2(2+\tilde{u}')}\approx\frac{1}{4},\hspace{1cm}
\tilde{u}_c^2\ll \frac{1}{4}
\end{equation}
The integral in Eq.(\ref{dimless-int}) in the interval $[0,\tilde{x}_c]$ can 
therefore safely be approximated by 
\begin{equation}
\label{dimless-int-xc}
\int\limits_{0}^{\tilde{x_c}} \mathrm{d}\tilde{x} 
\left[\tilde{u}^2+\frac{1}{2(2+\tilde{u}')} \right] = \frac{1}{4}\tilde{x}_c.
\end{equation}  
In the interval $[\tilde{x}_c,\tilde{D}-\tilde{x}_0]$ the approximation for the splay, 
Eq.(\ref{u-approx}), is valid. 
The integral in Eq.(\ref{dimless-int}) can then be 
rewritten in the following form
\begin{eqnarray}
\label{dimless-int-rest}
\int\limits_{\tilde{x}_c}^{\tilde{D}-\tilde{x}_o} \mathrm{d}\tilde{x} \left[ \tilde{u}^2+ 
\frac{1}{2(2+\tilde{u}')} \right]\nonumber\\
= \frac{1}{2}\int\limits_{\tilde{x}_c}^{\tilde{D}-\tilde{x}_o}\mathrm{d}\tilde{x} \left[\tilde{u}\tilde{u}'
-\tilde{u}^2\tilde{u}'-\frac{\tilde{u}'}{4}+\frac{1}{2}\right]\nonumber\\ 
= \left[\frac{\tilde{x}}{4}-\frac{\tilde{u}}{8}+\frac{\tilde{u}^2}{4}-\frac{\tilde{u}^3}{6}
\right]_{\tilde{x}_c}^{\tilde{D}-\tilde{x}_o}.
\end{eqnarray}
By construction $\tilde{u}(\tilde{x}_c)=\tilde{u}_c \ll 1$, see Eq.(\ref{u-approx}). 
Therefore the integral in 
Eq.(\ref{dimless-int}) in the limits $[0,\tilde{D}-\tilde{x}_0]$ can be very well approximated by making 
use of Eqs.(\ref{dimless-int-xc},\ref{dimless-int-rest}):
\begin{equation}
\int\limits_{0}^{\tilde{D}-\tilde{x}_o} \mathrm{d}\tilde{x} \left[ \tilde{u}^2+ 
\frac{1}{2(2+\tilde{u}')} \right] 
= \frac{\tilde{D}-\tilde{x}_0}{4}-\frac{\tilde{u}_D}{8}+\frac{\tilde{u}_D^2}{4}-\frac{\tilde{u}_D^3}{6}.
\end{equation}
So far we calculated only one part of the brush. The one from the ``zero splay chain'' to the open end. 
In Fig.(\ref{const-fig}) this is the right part from $0$ to $D-x_0$. The left part from $0$ to $x_0$ or 
rather from the ``zero splay chain'' to the surface can be calculated completely analogous 
replacing $\tilde{D}-\tilde{x}_0$ with $\tilde{x}_0$ and $\tilde{u}_D$ with $\tilde{u}_0$.   
Adding up the results for both parts in both intervals, accounting for the prefactors in Eq.(\ref{Fbrush}) 
and converting back to variables carrying dimensions 
we get as a final result for the free energy of the chains forming the finite brush
\begin{eqnarray}
\label{FbrushI}
\beta F_{brush} &=& \frac{d}{N b^2}\sigma \left[3Dh^2-h^2(u_0+u_D)\right.\nonumber\\
&+&\left.\frac{h}{\sqrt{2}}(u_0^2+u_D^2)
-\frac{1}{6}(u_0^3+u_D^3)\right].
\end{eqnarray}
Plugging in the $\alpha$-dependent expressions for $\sigma$, $D$, $u_0$ and $u_D$ and adding $F_{rod}$ 
we get the $\alpha$-dependent part of the total free energy of the system.
\begin{eqnarray}
\label{Ftot}
\beta F(\alpha) = 2 f \beta \gamma d^2 \tan(\alpha)+\frac{3}{4^{2/3}} f N \left(\frac{b}{d}\right)^{4/3} \cos(\alpha)^{2/3}\nonumber\\
+ \frac{N^2}{4} \left(\frac{b}{d}\right)^3 \cos^2(\alpha)\times\nonumber\\
\left[\frac{1}{\sqrt{2}}-\frac{7}{6}-\tan(\alpha)+\frac{\tan^2(\alpha)}{\sqrt{2}}-\frac{\tan^3(\alpha)}{6}\right]\nonumber\\
\end{eqnarray}
This equation is only valid for $0 \le \alpha \le \pi/4$, since $u_0(\pi/4) = h = u_D$. In the interval 
$\pi/4 < \alpha < \pi/2$ the splay at the surface remains constant at its maximum value $u_0 = h$, 
and the term in square brackets in Eq.(\ref{Ftot}) reduces to $[\sqrt{2}-7/3]$. 

There are two regimes in each of which the above free energy, Eq.(\ref{Ftot}),
shows a different behaviour. The regime where the chains are rather short
shows a first order like transition. With decreasing $\gamma$ there is a jump
from a stable phase with no shift ($\alpha=0$) to a stable phase with a large
shift ($\alpha \gg 0$). The other regime in which the chains are rather long
shows a second order transition from a stable phase with $\alpha=0$ to a phase
with finite $\alpha$.

\begin{figure}[h]
\includegraphics[width=\linewidth]{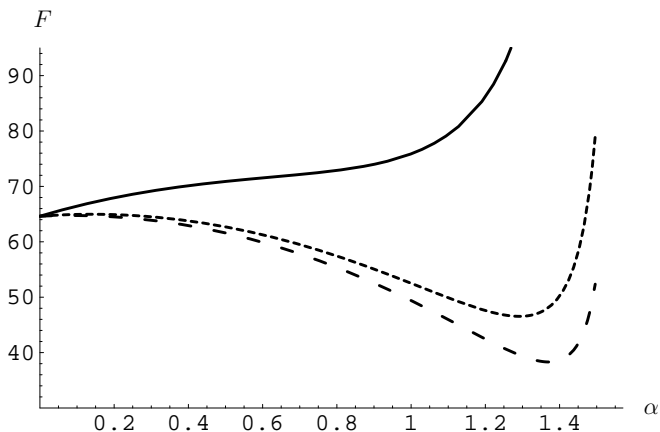}
\caption{\label{FalphaI-fig}This plot shows the dependence of the free energy in units of $kT$ on $\alpha$ 
for different values of $\gamma$ in the regime $N<N_t$. Parameters: $f = 10,  b/d = 0.2,  N = 50,
\beta\gamma d^2 = 1$ (upper curve),  0.25,  0.15 (lower curve). 
In this regime the effect of decreasing grafting density with increasing $\alpha$ 
dominates the effect of the splay of the coils close to the surface.}
\end{figure}

As a criterion to distinguished these two regimes the chain length can be
used. The chain length $N_t$ which separates these two regimes is given by
$N_t \approx 1.4 f \left(\frac{d}{b}\right)^{5/3}$.  For $N<N_t$ the
transition is first order like and continuous for $N>N_t$. This condition can
be interpreted such that for $N<N_t$ the second term in the free energy
(Eq.(\ref{Ftot})) dominates the third term. The second term solely represents
the effect of decreasing grafting density, whereas the third term also
represents the effect of increasing splay of the chains close to surface.
Therefore, if the grafting density effect is dominant, the transition is
similar to the tilting transition observed in lamellar structures of rod-coil
copolymers, see e.g. \cite{halperin1}. Since then the minimum in the free
energy is essentially given by the balance of $\tan{\alpha}$ (first term) and
$\cos{\alpha}$ (second term) there cannot be a continuous transition with a
minimum at small values of $\alpha$.  To illustrate this behaviour, the free
energy as a function of $\alpha$ for different values of $\gamma$ is plotted
in Fig.(\ref{FalphaI-fig}).

\begin{figure}[h!]
\includegraphics[width=\linewidth]{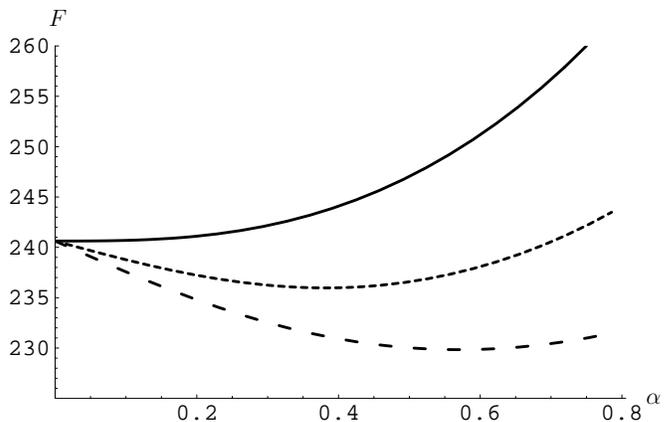}
\caption{\label{FalphaII-fig} This plot shows the dependence 
  of the free energy in units of $kT$ on $\alpha$ for different values of
  $\gamma$ in the regime $N>N_t$. Parameters: $f = 10, b/d = 0.2, N = 200,
  \beta\gamma d^2 = 4$ (upper curve), 3, 2.5 (lower curve).  In this regime the
  effect of increasing splay of the chains close to the surface determines the
  behaviour of the free energy.}
\end{figure}

For $N \ge N_t$ the third term in the free energy, Eq.(\ref{Ftot}), gets equal
to or bigger than the second term.  This means that the increase in splay of
the chains close to the surface becomes important. Since $u_0$ scales with
$\tan{\alpha}$ like the contribution of the rods ($F_{rod}$) does, a
continuous transition is now possible.  The dependence of the free energy on
$\alpha$ for different values of $\gamma$ is shown in
Fig.(\ref{FalphaII-fig}).

The assumption of the shift to be constant is an oversimplification. The
crossover from one regime ($N \ge N_t$) in which a shift develops continuously
to a regime ($N<N_t$) in which the system shows a first order transition like
jump from zero shift to large finite shift is an unphysical artefact of this
approximation. Only in the limit of a very large number of copolymers $f$
forming one lamellar like aggregate the constant shift assumption might be
reasonable. However, in this limit the effect of the surface becomes
negligible and the system always shows a tilting transition - compare
\cite{halperin1} - even if not in contact with the surface.
\end{appendix}

\end{document}